# Dynamic Power Management in Modular Reconfigurable Battery Systems with Energy and Power Modules

Pouyan Pourhadi, Nima Tashakor, Mahdi Bayati, and Stefan Goetz

*Abstract*—Integrating power electronics with batteries can offer many advantages, including load sharing and balancing with parallel connectivity. However, parallel batteries with differing voltages and power profiles can cause large circulating currents and uncontrolled energy transfers, risking system instability. To overcome these challenges, we propose a novel modular reconfigurable topology for AC batteries, employing coupled inductors between adjacent submodules. This approach dynamically manages energy distribution between energy and power modules. Under normal conditions, energy modules operate in series or parallel configurations to deliver stable voltage and current, ensuring efficient power delivery. During heavy loads, such as during electric-vehicle acceleration, power modules seamlessly meet additional power demands beyond the energy modules' capacity. The use of mutual inductors reduces inductance along the load path for faster response times while providing sufficient inductance to regulate circulating currents and enable efficient energy transfer among modules. The system's AC load further complicates control, necessitating a simple yet effective feedforward-feedback control strategy to maintain satisfactory performance. The experimental results confirm the system's reliability and effectiveness, achieving over 92% power efficiency under dynamic load changes. Steady-state conditions demonstrate balanced current sharing, while the advanced coupled inductor design and optimized control strategy reduce core size by 90% and switching losses by 15%, significantly enhancing overall efficiency.

## I. Introduction

Cascaded bridge (CB) and modular multilevel converters (MMCs) have secured a key position in power electronics as they offer superior power quality and reduce filter requirements [1]-[2]. However, existing cascaded bridge designs are limited by their basic series and bypass configurations, which restrict their flexibility and efficiency in dynamic applications such as electric vehicles [3]-[5]. The integration of batteries with different voltages, type, or states of charge further complicates operation and leads to challenges such as uncontrolled energy transfer, circulating currents, and instability, especially under rapid load changes and high-power demand [6]-[9]. These problems require precise sensing and control to maintain voltage balance [10]-[18].

Advancements in MMC and CB topologies, including diode/switch-clamped [19]-[24], series/parallel [25]-[30], and energy-exchanging designs [31]-[35], have highlighted the potential of reconfigurable battery systems. These innovations can easily balance module voltages and employ modulation strategies to manage energy exchange separately from the main load [36]-[38]. However, limitations in load-sharing and varying semiconductor current ratings reduce efficiency and suppress battery ripple currents [39]-[41].

These recent advancements have been applied to batteries and enabled reconfigurable battery management by segmenting the system voltage into discrete module voltages to support module states such as series, parallel, and bypass. Parallel connectivity between modules introduces unique opportunities. It facilitates load sharing and energy balancing [42]. However, uncontrolled current flow in case of systematic voltage disparities can lead to significant power losses and thermal stress on components. Due to the large heterogeneity of even same-time batteries after various years of service, second-life batteries have a low remaining value and require substantial effort for characterization and refurbishing with conventional battery system technology so that they cannot compete with new batteries. Moreover, differences in battery chemistry result in distinct cell voltages and voltage profiles, charge-discharge rates, and thermal behavior [41]-[51]. System integration of such heterogeneous batteries

Reconfigurable battery architectures dynamically rewire a battery and adjust the energy flow path based on internal and external objectives and constraints. Those objectives and constraints include demand, energy and conversion losses, balancing, and battery operation. Reconfigurable battery systems promise the combination of energy and power modules each to serve distinct but complementary roles for higher system-level performance and adaptability [52]-[53]. Energy modules could primarily provide steady-state power delivery and operate in either series or parallel configuration, which readily maintains voltage balance across the system. These modules offer high energy density for applications requiring prolonged energy output. Conversely, power modules can provide rapid high-power bursts during events for stabilizing a grid or to accelerate and break a car. These modules are characterized by their higher power density and ability to deliver short-term energy at a high rate. The integration of these modules within a reconfigurable framework promises seamless energy redistribution based on demand and enables the system to adapt dynamically to varying load profiles [54]-[61].

To solve the problems of different batteries and battery chemistries in one system, this work proposes a modular reconfigurable AC battery system with magnetic inter-module connections for seamless bidirectional energy transfer between modules. The novel topology dynamically allocates power between energy and power modules based on load conditions to optimize system performance, while it reduces magnetics size and

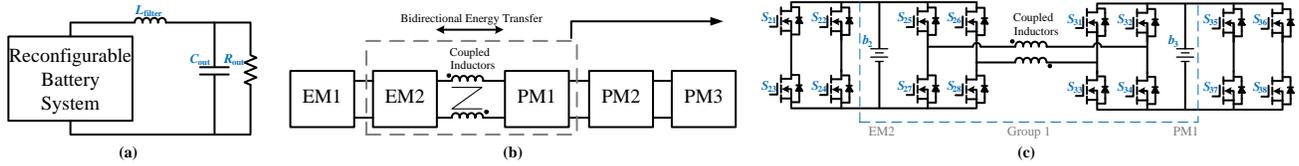

Fig. 1 a) Reconfigurable battery system circuit. (b) Internal structure with bidirectional energy transfer via coupled inductors. (c) Energy and power modules with coupled inductor configuration.

switching losses. Coupled inductors enable efficient bidirectional energy transfer between the modules and ensure that the additional power required during load spikes is effectively supplied. This bidirectional transfer capability allows for rapid response to load changes without compromising system stability and efficiency. The inter-module DC–DC conversion functionality can manage power exchange not only during steady-state operation but also during dynamic changes in load conditions. Coupled inductors enable minimal inductance along the load current path for rapid response, whereas they on the other hand provide sufficient inductance for circulating currents and controlled efficient module-to-module energy transfer.

The rest of the paper is organized as follows. Section II, III, IV, V, and VI present the proposed topology's operation principles, design considerations, and comparison with recent work, along with simulations and experiments to verify the analysis and evaluate system performance.

## II. Topology Description

Figure 1 shows the proposed power circuit consists of five modular units. Each unit features asymmetrical double-full-bridge submodules connected in parallel. Coupled inductors are strategically placed between specific submodules (e.g., between Modules 2 and 3) to link energy and power modules, each with different battery voltages. Each group includes the right full-bridge of Module j, the left full-bridge of Module (j+1), their batteries, and interconnecting components such as coupled inductors.

The design integrates inter-module DC–DC conversion functionality directly with the submodule structure. The transistors in each module perform dual roles:
1. Controlling the load current ($i_{out}$) for efficient power delivery.
2. Enabling bidirectional buck-boost DC–DC energy transfer between adjacent modules.

The coupled inductors have two key features:
1. Common-mode balancing: The coupled inductors evenly split $i_{out}$ to reduce imbalances during steady-state operation.
2. Differential-mode operation: Substantial inductance in the differential mode enables controlled energy transfer between modules with different battery voltages.

This architecture enables seamless energy redistribution, allowing the system to adapt dynamically to fluctuating load conditions while maintaining balanced performance across modules.

## III. Operation Principle

### A. Operation Modes of the Interconnections Between the Energy and Power Modules

Figure 2(a) illustrates how the system functions in four distinct operation modes: series, parallel, buck/boost, and bypass. These modes enable dynamic energy transfer and voltage balancing while adapting to varying power demands. The use of coupled inductors ensures effective current sharing and energy management across modules.

### Mode 1: Series Connection and Differential Choke

As Fig. 2(a) shows, when $S_{27}$, $S_{28}$, $S_{31}$, and $S_{32}$ are active, a series connection between the two modules is established. The inductor current comprises two components: one that forms $i_{out}$ and the other the circulating current ($i_{circ}$). To solve current imbalances due to battery voltage differences, a differential inductor bridges the second double full bridge submodule to the third one. This modification introduces an inherent DC-DC stage for adaptive battery voltage adjustment and can control the power and energy contributions of modules, which will be explained later.

The currents $i_{out}$ and $i_{circ}$ determine the inductor currents according to
$$i_1 = \alpha \cdot i_{out} + i_{circ}, \quad i_2 = \alpha \cdot i_{out} - i_{circ}, \tag{1}$$
where $\alpha$ represents the current sharing ratio between the inductors.

### B. Output Voltage Calculation and Effective Inductances

The number of modules connected in series determines the string's output voltage ($v_{out}$), expressed as
$$v_{out} = \sum_{x=i} (s_i \times v_{bi}), \tag{2}$$
where $s_i$ can be either 0 (for parallel or boost/buck modes) or 1 (for series mode). The module voltage $v_{bi}$ represents the battery voltage. This equation highlights that both energy and power modules contribute to the output voltage. The contribution of all modules increases the flexibility compared to topologies where batteries are simply connected in parallel or are even bypassed. Figure 2(b) shows the equivalent circuit in the series mode.

According to Fig. 2(b), $i_{out}$ is the summation of $i_1$ and $i_2$. To analyze the inductance effects, the effective common-mode and differential-mode inductances are derived as
$$L_{eff,CM} = 0.5(L_1 - M_{12}) + \Delta r, \tag{3}$$
$$L_{eff,DM} = L_1 + M_{12} + L_2 + M_{21} + \Delta L, \tag{4}$$
where $\Delta r$ and $\Delta L$ account for additional resistive and inductive offsets introduced by design or circuit variations. $L_1$, $L_2$, $M_{12}$, and $M_{21}$ are self and mutual inductances. These adjustments enhance the accuracy of energy transfer calculations and stabilize $i_{circ}$.

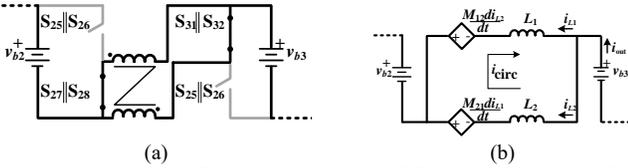

Fig. 2 (a) Serial mode of asymmetrical double-full-bridge modules and (b) the equivalent circuit of serial mode.

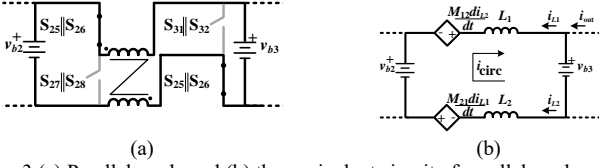

Fig. 3 (a) Parallel mode and (b) the equivalent circuit of parallel mode.

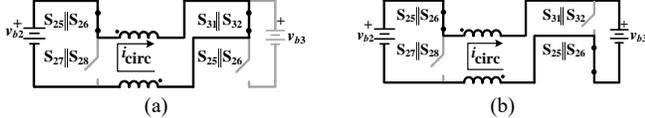

Fig. 4 Buck mode (left to right) (a) linearly charging the inductors, (b) turning to parallel mode.

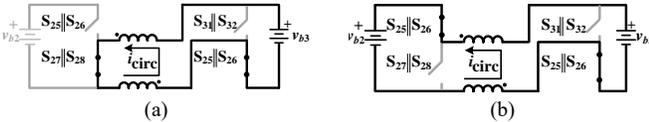

Fig. 5 Boost mode (right to left) (a) linearly charging the inductors, (b) turning to parallel mode.

The dynamic equation governing $i_\text{circ}$ in series mode is derived from Kirchhoff's Voltage Law as

$$L_\text{eff,DM}\frac{d}{dt}i_\text{circ} + L_\text{eff\_CM}\frac{d}{dt}\alpha i_\text{out} = 0. \quad (5)$$

Given the high differential-mode inductance, $(d/dt\, i_\text{circ}) \approx 0$, which indicates a quasi-static circulating current in series mode.

### Mode 2: Parallel Mode and Voltage Difference

In the parallel mode (Fig. 3(a)), the switches of the series mode are toggled: switches $S_{25}$ and $S_{33}$ are turned on and switches $S_{31}$ and $S_{27}$ are turned off. This configuration splits the common-mode current ($i_\text{out}$) equally between the inductors. The rate of change of $i_\text{circ}$ depends on the voltage differences between the batteries. Figure 3(b) shows the equivalent circuit for the parallel mode and entails

$$v_{b3} - v_{b2} = (L_1 + M_{12} + M_{21} + L_2 + \Delta L)\frac{di_\text{circ}}{dt} + \Delta V_\text{offset}, \quad (6)$$

where $\Delta V_\text{offset}$ accounts for minor asymmetries in module voltages or inductance mismatches.

### Mode 3: Buck and Boost Modes

By creating a slight shift in the switch toggling instances of the two full-bridges, two additional modes, known as buck and boost modes can emerge (Figs. 4 and 5). Suppose $v_{b2} > v_{b3}$ similar to the conventional buck converter. When $S_{25}$, $S_{33}$, and $S_{28}$ are already active, $S_{32}$ will be turned off and $S_{34}$ will be turned on, the left battery magnetizes the coupled inductors to that $i_\text{circ}$ increases in a clockwise direction (Fig. 4(a)) per

$$\frac{d}{dt}i_\text{circ} = \frac{2m_d v_{b2}}{L_\text{eff,DM}}, \quad (7)$$

where $2m_d$ is the duty cycle controlling the buck operation. The stored energy is then discharged into the lower-voltage module during the parallel mode.

In a specified period, $v_{b2}$ is used to charge the interconnected inductors. Next, $S_{34}$ and $S_{32}$ are adjusted to transition the system into the parallel mode (Fig. 4(b)), where $S_{32}$ closes and $S_{34}$ opens. According to (6), the energy stored in the interconnected inductors is then transmitted to $v_{b3}$ via a circulating current. Also, in Fig. 5(a) when $S_{32}$, $S_{33}$ and $S_{28}$ are already active, $S_{25}$ will be turned off and $S_{27}$ will be turned on, and then, $i_\text{circ}$ in the anticlockwise direction is increased by using $v_{b3}$. This case is similar to the conventional boost converter, which consists of an inductor, a switch, a diode, and a capacitor (in this study, battery). According to Fig. 5(b), after charging the coupled inductors, $S_{27}$ is opened and $S_{25}$ closed, which turns the interconnection to the parallel mode and transfers the energy of the coupled inductors into $v_{b2}$ in the form of a circulating current per (6). The ratio between the boost/buck mode and the parallel mode controls $i_\text{circ}$ in both directions without affecting $i_\text{out}$.

### Mode 4: Bypass Mode

The bypass mode is only invoked when a battery module's participation in output voltage generation is not required. This is achieved by activating either all top or alternatively all bottom switches.

### B. Voltage Balancing and Inductor-Based Energy Transfer Control

Efficient voltage balancing is achieved by alternating between the parallel and the boost/buck mode. The voltage difference across modules determines the circulating current [62], [63] per

$$\frac{d}{dt}i_\text{circ} = \frac{v_{bj} - v_{b(j+1)}}{L_\text{eff,DM}} + \frac{d}{dt}\left(\frac{1}{C_\text{inductor}}\int i_\text{circ} dt\right). \quad (8)$$

This formulation ensures that voltage discrepancies decay exponentially over time and stabilizes the system. The inductance $L_\text{eff,DM}$ is chosen such as to balance speed and stability with its time constant $T$ defined as

$$T = \frac{L_\text{eff,DM}}{R}. \quad (9)$$

For fast energy balancing, the inductance is typically in the micro-henry range, which allows quick adjustments without excessive current ripple.

### C. Switching Logic

The switching logic used in this system is based on phase-shifted carrier (PSC) modulation and follows

$$\text{state } L_{k(k\in\mathbb{N})} = \begin{cases} \text{State 1: } \begin{bmatrix} 0 & 0 \\ 1 & 1 \end{bmatrix}, & m_0 - m_{d2} > C_k, \\ \text{State 0: } \begin{bmatrix} 1 & 0 \\ 0 & 1 \end{bmatrix}, & m_0 - m_{d2} < C_k, \end{cases} \quad (10)$$

$$\text{state } U_{k(k\in\mathbb{N})} = \begin{cases} \text{State 1: } \begin{bmatrix} 1 & 1 \\ 0 & 0 \end{bmatrix}, & m_0 + m_{d2} > C_k, \\ \text{State 0: } \begin{bmatrix} 0 & 1 \\ 1 & 0 \end{bmatrix}, & m_0 + m_{d2} < C_k. \end{cases} \quad (11)$$

Each carrier $C_k$ serves alongside the switch states of the right ($S_{k5}$, $S_{k6}$, $S_{k7}$, and $S_{k8}$) and left ($S_{(k+1)1}$, $S_{(k+1)2}$, $S_{(k+1)3}$, and $S_{(k+1)4}$) full bridges of the $K$-th group, denoted as $L_k$ and $U_k$. The modulation indices $m_0$ and $m_{d2}$ respectively govern output voltage and energy transfer control. The controller can set negative and positive $m_{d2}$ values.

### A. 1: Switching Modes in Series, Parallel, and Boost

Fig. 6 portrays the full-bridge states with two different modulation indices as an example. Series connection occurs when both full bridges are in State 1, which maintains a constant circulating current. Conversely, a parallel connection arises when both modules are in State 0 so that a circulating current to fluctuate as per (6), and depending on the modulation, from higher-voltage to lower-voltage modules or vice versa. In cases where $L_k$ is in State 0 and $U_k$ is in State 1 ($m_{d2}>0$), the circulating current grows clockwise. Conversely, $L_k$ in State 1 and $U_k$ in State 0 entails anticlockwise circulating current ($m_{d2}<0$). The durations for series, parallel, and boost states are

$$\begin{cases} T_s = T_{sw}(m_0 - m_{d2}), & \text{Series state,} \\ T_p = T_{sw}(1 - m_0 - m_{d2}), & \text{Parallel state,} \\ T_b = 2T_{sw}m_{d2}, & \text{Boost state.} \end{cases} \quad (12)$$

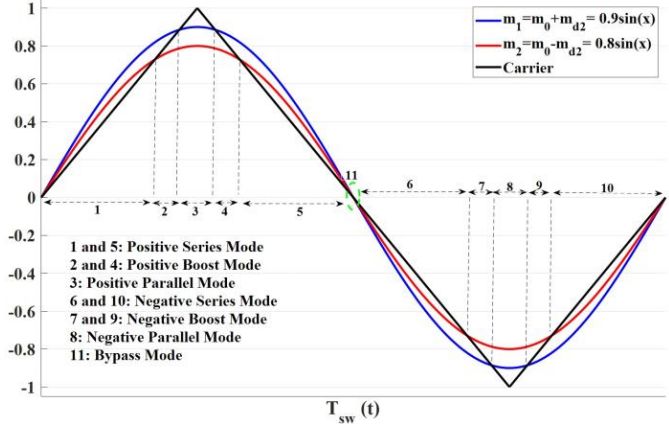

Fig. 6 The full-bridge states ($L_k$ and $U_k$).

### A. 2: Control Strategy and Diagram

Figure 7 illustrates the proposed system's control diagram. The control approach is split into two relatively independent components: one for regulating output voltage, employing a straightforward PI controller (PI$_1$) with a feedback loop to compute $m_0$. Additionally, a second PI controller (PI$_2$) manages energy transfer between modules and computes an $m_{d2}$ for each controllable interconnection.

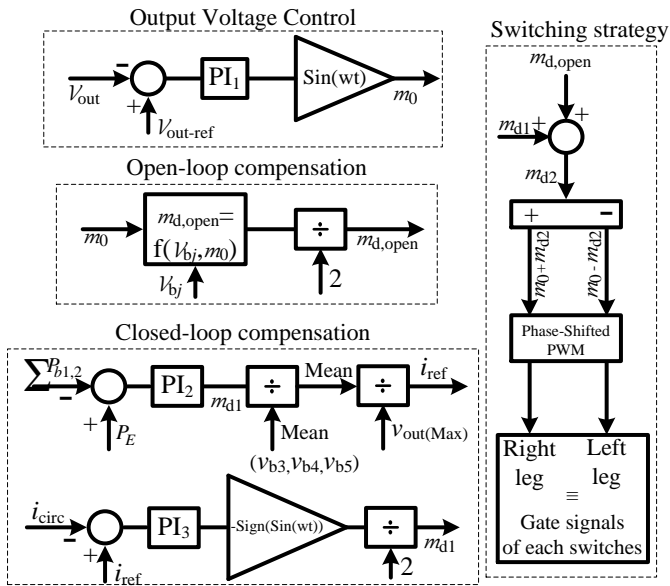

Fig. 7 Control System of the modulation of the module groups.

### A. 3: Bypass Mode Criteria

The bypass mode is activated when the absolute value of $2m_{d2}$ is larger than both $m_1$ and $m_2$. This happens when the modulation indices cross zero, which prevents the coupled inductor from sudden changes in the circulating current (indicated with Reference Number 11 in Fig. 6)

### C. Energy transfer between the modules consisting of the coupled inductor

In this full-bridge configuration, the steady-state circulating current ($i_{circ,SS}$), which results from the voltage difference between adjacent modules, follows

$$i_{circ,SS} = \frac{v_{bj} - v_{b(j+1)} + K_{dynamic} \cdot \Delta SOC_{j+1}}{R_{eq} + Z_{dyn}}, \quad (13)$$

where $R_{eq} = 4R_{DS(on)}(T) + R_{b_j} + R_{b_{(j+1)}} + R_L$, includes contributions from the on-resistance of MOSFET switches (temperature-dependent), internal battery resistances, and the equivalent series resistance (ESR) of the inductor. $Z_{dyn}$ accounts for additional dynamic impedance contributions. $K_{dynamic}$ reflects the impact of state-of-charge (SOC) differences on the circulating current. $\Delta SOC_{j+1}$ represents the SOC imbalance between modules. This circulating current flows from Module $j$ to Module $(j+1)$ in parallel mode if $v_{b_j}$ is greater than $v_{b_{j+1}}$.

To ensure no net flux accumulation in the coupled inductors, the voltage integration across the inductors in each mode must average to zero over a switching period. Neglecting resistive elements in series mode, the voltage integrals are as follows: for the parallel mode

$$\int v_{Lp} \, dt = (v_{b_j} - v_{b_{j+1}} - i_{circ} K_R) T_p, \quad (14)$$

where $K_R$ accounts for the resistive effects of the path; for the boost/buck mode

$$\int v_{Lb} dt = (-v_{b_{j+1}} - i_{circ} K_B) T_b, \quad (15)$$

where $K_B$ accounts for dynamic resistance and switching effects.

The integrals incorporate both the voltage difference between modules and the effect of the equivalent resistance, $R_{eq}$. For balanced operation without flux accumulation, the average voltage across the inductors over a switching period must be zero, which leads to the relationship between $T_p$ and $T_b$ of

$$\frac{v_{b_j} - v_{b_{j+1}}}{v_{b_{j+1}}} = \frac{T_b}{T_p}. \quad (16)$$

By substituting for $T_p$ and $T_b$ in terms of modulation indices, the expression becomes

$$\frac{v_{b_j} - v_{b_{j+1}}}{v_{b_{j+1}}} = \frac{2m_{d2}}{(1 - m_0 - m_{d2})}. \quad (17)$$

This relationship yields the open-loop modulation index $m_{d,open}$ required to maintain a circulating current near zero. Therefore, $m_{d,open}$ for the Interconnection Group $j$ is given by

$$m_{d,open} = \frac{(v_{b_j} - v_{b_{j+1}})(1 - m_0)}{v_{b_j} + v_{b_{j+1}}}. \quad (18)$$

If $v_{b_j} < v_{b_{j+1}}$, $m_{d,open}$ will be negative, which indicates reverse energy flow.

To achieve a non-zero circulating current ($i_{\text{circ,SS}}$), the additional voltage drop across the equivalent resistance $R_{\text{eq}}$ must be considered. The voltage across the inductors, accounting for $i_{\text{circ}}$, is expressed as

$$\int v_L \, dt = \begin{cases} \left(v_{b_j} - v_{b_{j+1}} - i_{\text{circ}} R_{\text{eq}}\right) T_p, & \text{parallel state,} \\ \left(-v_{b_{j+1}} - i_{\text{circ}} R_{\text{eq}}\right) T_b, & \text{boost/buck state.} \end{cases} \quad (19)$$

Substituting these expressions and solving for $m_{d,\text{open}}$ leads to

$$m_{d,\text{open}} = \frac{\left(v_{b_j} - v_{b_{j+1}} - i_{\text{circ}} R_{\text{eq}}\right)(1 - m_0)}{v_{b_j} + v_{b_{j+1}} + i_{\text{circ}} R_{\text{eq}}}. \quad (20)$$

This open-loop value of $m_{d,\text{open}}$ acts as a feed-forward control term that adjusts the output based on the desired circulating current, voltage difference, and resistive losses. The feed-forward part enables fast and precise management of energy flow between modules.

To support dynamic control objectives, the reference circulating current $i_{\text{circ,ref}}(t)$ can be adjusted for tasks such as charge balancing or load compensation per

$$i_{\text{circ,ref}}(t) = K_{\text{SOC}}\left(SOC_{\text{avg}} - SOC_{j+1}\right) + K_{\text{load}} i_{\text{load}}, \quad (21)$$

where $K_{\text{SOC}}$ is a gain for adjusting $i_{\text{circ,ref}}$ based on state of charge (SOC) and $K_{\text{load}}$ accounts for load current compensation and ensures a balanced load distribution. This approach allows the system to dynamically adapt $i_{\text{circ}}$ for real-time conditions to improve both energy balancing and stability. Setting $i_{\text{circ}} = 0$ minimizes module-to-module energy exchange, while non-zero $i_{\text{circ}}$ values help balance $SOC$ across modules per

$$i_{\text{circ},j} = SOC_{\text{avg}} - SOC_{j+1}. \quad (22)$$

$i_{\text{circ},j}$ alternates its direction within each half switching cycle due to the inherent periodic nature of the switching operation. This alternating behavior is fundamental to the system's design and ensures a bidirectional energy flow between adjacent modules. The bidirectional behavior ensures efficient energy redistribution between modules. Mathematically, this alternation is expressed as

$$i_{\text{circ},j}(t) = K_{\text{SOC}}\left(SOC_{\text{avg}} - SOC_{j+1}\right) \cdot \sin(wt), \quad (23)$$

where $\sin(wt)$ models the alternating behavior of the current within each switching cycle and $w$ represents the angular frequency of the switching signal.

The periodic sign change in $i_{\text{circ},j}$ facilitates:
1. SOC Balancing: Alternating directions progressively reduce SOC imbalances between adjacent modules.
2. Stability Enhancement: Oscillatory currents prevent energy buildup in any single module, avoiding overheating or inefficiencies.

The modulation indices ($m_{d,\text{open}}$) and duty cycles of the switching signals dynamically control the magnitude of $i_{\text{circ},j}$. For $SOC$ balancing, the reference circulating current $i_{\text{circ,ref}}(t)$ can be expressed as

$$i_{\text{circ,ref}}(t) = K_{\text{SOC}}\left(SOC_{\text{avg}} - SOC_{j+1}\right). \quad (24)$$

The alternating sign of $i_{\text{circ},j}$ ensures that the net energy transfer over a full switching period maintains energy balance across the modules. During the parallel mode, $i_{\text{circ},j}$ equalizes voltage differences, while in the series mode, periodic reversal stabilizes energy transfer and prevents flux accumulation in inductors, ensuring smooth operation across modes.

The proposed system achieves robust energy transfer management across full-bridge modules through both the open-loop modulation indices and dynamic circulating current adjustments.

## IV. CONVERTER DESIGN CONSIDERATIONS

To verify the proposed system's performance in the full-bridge configuration, it is essential to design the components for continuous-conduction mode (CCM) operation. The continuous conduction mode ensures steady power flow and minimizes switching transients.

The cross-sectional area $A_p$ of the magnetic core is a critical parameter for the inductor design, particularly to avoid saturation at the peak of the circulating current. For full-bridge systems, $A_p$ can be determined by

$$A_p = \left[\frac{K_i L I^2 \sqrt{1+\gamma}}{B_{\max} K_t \sqrt{K_u \Delta T}}\right]^{8/7}, \quad (25)$$

where $K_t$ is set to 48,200 (a constant related to core characteristics) [64] and $I$ is the current, proportional to the maximum amplitude of the circulating current. Assuming constant values of $\gamma$, $B_{\max}$, $\Delta T$, $K_i$, and $K_u$, $A_p$ is proportional to $I^{16/7}$, which for the proposed topology is equal to the maximum amplitude of the circulating current. For the full-bridge configuration, $A_p$ should be sized to handle higher peak currents than a half-bridge, given that the circulating current will be doubled due to the full-bridge architecture. This choice ensures that the inductor operates without saturating at peak currents, even under load transients.

Assuming continuous conduction and that $v_{b_j} > v_{b_{j+1}}$, the inductor current, $i_L^{\text{magnetize}}$ fulfills

$$i_L^{\text{magnetize}} \leq i_{\text{circ}} \leq i_{\text{circ}}^{\text{rated}} + \frac{2 v_{b_{j+1}} m_{d2}}{L_{\text{eff,DM}} f_{\text{sw}}}. \quad (26)$$

To limit the peak local (magnetizing) circulating current, the required inductor value, $L_{\text{eff,DM}}$ should satisfy

$$L_{\text{eff,DM}} \geq \frac{2 v_b^{\max} m_{d2}}{\Delta i_{\text{diff}} f_{\text{sw}}}, \quad (27)$$

where $i_{\text{diff}} = i_{\text{out}}^{\text{rated}} - i_{\text{circ}}^{\text{rated}}$. $i_{\text{out}}^{\text{rated}}$ and $i_{\text{circ}}^{\text{rated}}$ are respectively the rated output and circulating currents,. This inductor sizing ensures the inductor can handle the peak circulating current without saturation. The value $L_{\text{eff,DM}}$ is also adapted to the increased current demands of the full-bridge configuration.

An additional low-pass LC filter is formed with $L_{\text{filter}}$ and $C_{\text{out}}$ to smooth out switching ripple. For a maximum allowable output current ripple of 15%, the required minimum inductance for $L_{\text{filter}}$ is

$$L_{\text{filter}} \geq \frac{v_b(m_0)}{0.15 i_{\text{out}}^{\text{rated}} N f_{\text{sw}}}, \quad (28)$$

where $N$ is the number of modules, $f_{\text{sw}}$ is the switching frequency, and $m_0$ is the modulation index. The lower limit of the inductance ensures that output current ripple remains within acceptable limits.

To maintain voltage stability at the output, the output capacitor $C_{out}$ is selected to limit the voltage ripple to 1% of the output voltage ($\Delta v_{out} \leq 0.01 V_{out}$). The required minimum capacitance can be determined as

$$C_{out} = \frac{v_{out} D}{R_{out} \Delta v_{out} f_{sw}}, \tag{29}$$

where $D$ is the duty cycle and $R_{out}$ is the load resistance.

## V. Comparison with Similar Converters

### A. Comparison and Loss Analysis of the Proposed Converter Topology

Table I presents a comparison between the proposed topology and prior designs. The key distinction lies in the current flow between adjacent modules and the modulation strategy. Figure 8 illustrates three different interconnection circuits for the modules of the topologies listed in Table I: Figs. 8(a), 8(b), and 8(c) depict the connections in the proposed circuit and the literature [69, 72], [65-68, 70-71, 15, 74], and [73]. According to Fig. 8(a), the current passing through the coupled inductors is $i_{out}/2 + i_{circ}$, while in Fig. 8(b), the magnetization current is ($i_{out}/2 + i_{circ}$). However, in the proposed topology, only $i_{circ}$ contributes to the magnetization current, meaning that $i_{out}$, typically much larger than $i_{circ}$, does not impact the core size, allowing for a significantly smaller magnetic core in the proposed design.

Figure 8(c) illustrates an alternative design with a separate current path for the circulating current as presented in the literature [73]. While $i_{circ}$ flows through the inductor in this case, the amplitude of $i_{circ}$ is notably higher than in the proposed solution, as the load is not automatically divided between the two parallel modules. As a result, the RMS current in the batteries is considerably larger, despite a similar average current. When comparing losses, although the inductor values in the literature are akin to the proposed system, the total transistor losses are higher.

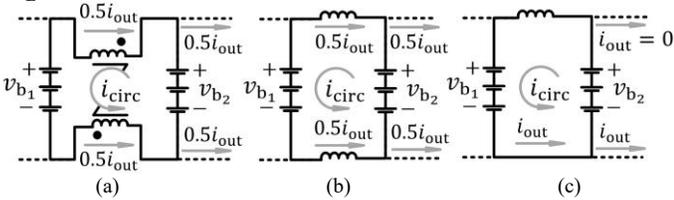

Fig. 8 $i_{circ}$ and $i_{out}$ flowing through the inductors in different topologies.

Configurations where all modules are active continuously [15], [65- 73, 74] cause higher losses due to continuous switching compared to the proposed system, which engages modules selectively.

Table I
Comparison with Prior Approaches

| Converter topology | Inductor current rating | Bidirectional Energy transfer | Load current sharing |
|---|---|---|---|
| [65] | $i_{out}/2 + i_{circ}$ | no | yes |
| [66] | $i_{out}/2 + i_{circ}$ | no | no |
| [67] | $i_{out}/2 + i_{circ}$ | no | no |
| [68] | $i_{out}/2 + i_{circ}$ | no | no |
| [69] | $i_{out}/2 + i_{circ}$ | no | yes |
| [70] | $i_{out}/2 + i_{circ}$ | no | no |
| [71] | $i_{out}/2 + i_{circ}$ | no | no |
| [72] | $i_{out}/2 + i_{circ}$ | no | yes |
| [73] | $i_{circ}$ | no | no |
| [15] | $i_{out}/2 + i_{circ}$ | yes | no |
| [74] | $i_{out}/2 + i_{circ}$ | no | no |
| Proposed converter | $i_{out}/2 + i_{circ}$ | yes | yes |

The proposed system reduces switching events by limiting active modules under normal conditions, leading to lower transistor losses and improved efficiency during typical operation.

As Table II demonstrates, the first two modules of the proposed circuit generate the output voltage, while the rest are connected in parallel and boost mode. In contrast to the previous literature [15, 65-74], however, all of the operational modes are active between the groups.

Switching and conduction losses ($P_{cond}$ and $P_{sw}$) in both the proposed system and the literature [69], [72], are calculated based on several parameters, including $i_{out}$ and $i_{circ}$. The transistor conduction losses are proportional to $(i_{out})^2$, and switching losses depend on $i_{out}$.

Table II
States between modules while energy modules are working

| Topology | [15], [65- 73, 74] | Proposed converter |
|---|---|---|
| Group 0: The left full bridge of module 1 and the right full bridge of module 5 | s, by | By |
| Group 1 | s, p, bo | s, p |
| Group 2 | s, p, bo | p, bo |
| Group 3 | s, p, bo | P |
| Group 4 | s, p, bo | P |

* Series: s, parallel: p, boost: bo, bypass: by

For the proposed converter, we must consider the negative half cycle of the modulation index. So, all of the $P_{cond_{Group\,0}}$ and $P_{sw_{Group\,0}}$ should be doubled:

$$P_{cond_{Group\,0}} = 2r_{DS}\left(\underbrace{2\left(\frac{(\frac{i_{out}}{2} + i_{circ})}{2}\right)^2 (1-m_0-m_{d2})}_{\propto P_{sw11}+P_{sw12}} + \underbrace{2\left(\frac{(\frac{i_{out}}{2} - i_{circ})}{2}\right)^2 (1-m_0-m_{d2})}_{\propto P_{sw55}+P_{sw56}}\right), \tag{30}$$

$$P_{\text{cond}_{\text{Group 1}}} =$$
$$2r_{\text{DS}} \left( \underbrace{2\left(\frac{(\frac{i_{\text{out}}}{2} + i_{\text{circ}})/2}\right)^2 (m_0 - m_{\text{d2}})}_{\propto P_{\text{sw17}} + P_{\text{sw18}}} + \underbrace{2\left(\frac{(\frac{i_{\text{out}}}{2} - i_{\text{circ}})/2}\right)^2 (m_0 + m_{\text{d2}})}_{\propto P_{\text{sw21}} + P_{\text{sw22}}} + \underbrace{\left(\frac{i_{\text{out}}}{2} + i_{\text{circ}}\right)^2 (1 - m_0 - m_{\text{d2}})}_{\propto P_{\text{sw15}}} + \underbrace{\left(\frac{i_{\text{out}}}{2} + i_{\text{circ}}\right)^2 (1 - m_0 - m_{\text{d2}})}_{\propto P_{\text{sw18}}} + \underbrace{\left(\frac{i_{\text{out}}}{2} - i_{\text{circ}}\right)^2 (1 - m_0 - m_{\text{d2}})}_{\propto P_{\text{sw22}}} + \underbrace{\left(\frac{i_{\text{out}}}{2} - i_{\text{circ}}\right)^2 (1 - m_0 - m_{\text{d2}})}_{\propto P_{\text{sw23}}} \right), \quad (31)$$

$$P_{\text{cond}_{\text{Group 2}}} = 2r_{\text{DS}} \left( \underbrace{\left(\frac{i_{\text{out}}}{2} + i_{\text{circ}}\right)^2 (1 - m_0 + m_{\text{d2}})}_{\propto P_{\text{sw25}}} + \underbrace{\left(\frac{i_{\text{out}}}{2} + i_{\text{circ}}\right)^2 (1 - m_0 + m_{\text{d2}})}_{\propto P_{\text{sw28}}} + \underbrace{\left(\frac{i_{\text{out}}}{2} - i_{\text{circ}}\right)^2 (1 - m_0 + m_{\text{d2}})}_{\propto P_{\text{sw33}}} + \underbrace{\left(\frac{i_{\text{out}}}{2} - i_{\text{circ}}\right)^2 (1 - m_0 - m_{\text{d2}})}_{\propto P_{\text{sw32}}} + \underbrace{\left(\frac{i_{\text{out}}}{2} - i_{\text{circ}}\right)^2 (2m_{\text{d2}})}_{\propto P_{\text{sw34}}} \right), \quad (32)$$

$$P_{\text{cond}_{\text{Group 3}}} = 2r_{\text{DS}} \left( \underbrace{\left(\frac{i_{\text{out}}}{2} + i_{\text{circ}}\right)^2 (1 - m_0 - m_{\text{d2}})}_{\propto P_{\text{sw35}}} + \underbrace{\left(\frac{i_{\text{out}}}{2} + i_{\text{circ}}\right)^2 (1 - m_0 - m_{\text{d2}})}_{\propto P_{\text{sw38}}} + \underbrace{\left(\frac{i_{\text{out}}}{2} - i_{\text{circ}}\right)^2 (1 - m_0 - m_{\text{d2}})}_{\propto P_{\text{sw42}}} + \underbrace{\left(\frac{i_{\text{out}}}{2} - i_{\text{circ}}\right)^2 (1 - m_0 - m_{\text{d2}})}_{\propto P_{\text{sw43}}} \right), \quad (33)$$

$$P_{\text{cond}_{\text{Group 4}}} = 2r_{\text{DS}} \left( \underbrace{\left(\frac{i_{\text{out}}}{2} + i_{\text{circ}}\right)^2 (1 - m_0 - m_{\text{d2}})}_{\propto P_{\text{sw45}}} + \underbrace{\left(\frac{i_{\text{out}}}{2} + i_{\text{circ}}\right)^2 (1 - m_0 - m_{\text{d2}})}_{\propto P_{\text{sw48}}} + \underbrace{\left(\frac{i_{\text{out}}}{2} - i_{\text{circ}}\right)^2 (1 - m_0 - m_{\text{d2}})}_{\propto P_{\text{sw52}}} + \underbrace{\left(\frac{i_{\text{out}}}{2} - i_{\text{circ}}\right)^2 (1 - m_0 - m_{\text{d2}})}_{\propto P_{\text{sw53}}} \right), \quad (34)$$

where $r_{\text{DS}}$ is the drain–source internal resistance. Equivalently, $P_{\text{cond}}$ for the topology [69], [72] follows

$$P_{\text{cond}_{\text{Group 0}}} =$$
$$2r_{\text{DS}} \left( \underbrace{2\left(\frac{(\frac{i_{\text{out}}}{2} - i_{\text{circ}})/2}\right)^2 (1 - 2m_{\text{d2}})}_{\propto P_{\text{sw11}} + P_{\text{sw12}}} + \underbrace{2\left(\frac{(\frac{i_{\text{out}}}{2} + i_{\text{circ}})/2}\right)^2 (1 - 2m_{\text{d2}})}_{\propto P_{\text{sw55}} + P_{\text{sw56}} + P_{\text{sw57}} + P_{\text{sw58}}} \right), \quad (35)$$

$$P_{\text{cond}_{\text{Group 1,2,3,4}}} = 2r_{\text{DS}} \left( \underbrace{\left(\frac{(\frac{i_{\text{out}}}{2} + i_{\text{circ}})/2}\right)^2 (1)}_{\propto P_{\text{sw18}}} + \underbrace{\left(\frac{(\frac{i_{\text{out}}}{2} - i_{\text{circ}})/2}\right)^2 (1 - 2m_{\text{d2}})}_{\propto P_{\text{sw22}}} + \underbrace{\left(\frac{(\frac{i_{\text{out}}}{2} + i_{\text{circ}})/2}\right)^2 (1 - m_0 + m_{\text{d2}})}_{\propto P_{\text{sw15}}} + \underbrace{\left(\frac{(\frac{i_{\text{out}}}{2} - i_{\text{circ}})/2}\right)^2 (1 - m_0 + m_{\text{d2}})}_{\propto P_{\text{sw23}}} + \underbrace{2\left(\frac{(\frac{i_{\text{out}}}{2} + i_{\text{circ}})/2}\right)^2 (m_0 - m_{\text{d2}})}_{\propto P_{\text{sw21}} + P_{\text{sw17}}} + \underbrace{\left(\frac{(\frac{i_{\text{out}}}{2} - i_{\text{circ}})/2}\right)^2 (2m_{\text{d2}})}_{\propto P_{\text{sw24}}} \right). \quad (36)$$

The switching losses for the proposed circuit and benchmark circuits in the literature [69], [72] follow

$$P_{\text{sw}_{\text{Group 0,1,2,3,4}}} = 5v_{\text{b}} \left( \underbrace{\frac{4\left(\frac{i_{\text{out}}}{2} - i_{\text{circ}}\right)/2}{}}_{\substack{\propto P_{\text{sw11}} + P_{\text{sw22}} \\ + P_{\text{sw13}} + P_{\text{sw14}}}} + \underbrace{\frac{4\left(\frac{i_{\text{out}}}{2} + i_{\text{circ}}\right)/2}{}}_{\substack{\propto P_{\text{sw57}} + P_{\text{sw58}} \\ + P_{\text{sw55}} + P_{\text{sw56}}}} \right) \frac{f_{\text{sw}}}{2} (t_{\text{ri}} + t_{\text{fi}} + t_{\text{rv}} + t_{\text{fv}}). \quad (37)$$

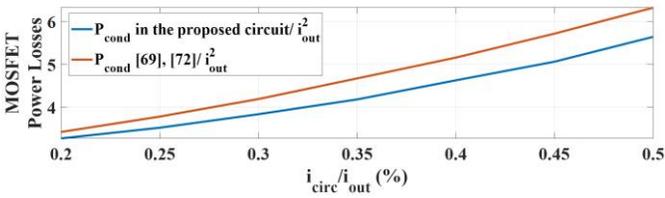

Fig. 9 Comparison between transistor conduction losses in [69], [72] and the presented circuit according to the ratio of $i_{circ}/i_{out}$.

This comparison considers batteries and inductors to be ideal. The losses are compared with the assumption of an equal effective modulation index, switching pattern, $r_{DS}$, $f_{sw}$, $v_b$, $t_{ri}$, $t_{fi}$, $t_{rv}$, and $t_{fv}$ for all switches. Consequently, $i_{out}$ and $i_{circ}$ are the only two effective factors in the losses. Figure 9 compares the transistor power losses of benchmark circuits from the literature [69], [72] with the presented work according to the ratio $i_{circ}/i_{out}$ based on (30)–(37). For all the ratios of $i_{circ}/i_{out}$, the total losses of the switches in the literature [69], [72] are much higher than the presented work.

Also, as previously stated, contrary to the literature [15], [65-68, 70-71, 74], the magnetization current in the proposed topology is only $i_{circ}$, decreasing the core size significantly. As a result, under this circumstance, the inductor current in the topologies in Figs. 8(a), (b), and (c) is either

$i_{out}/2 + i_{circ}$ [73] or
$i_{circ}$ [15], [69, 72], [65-68, 70-71, 74] and the proposed circuit. (38)

Additionally, according to (25), $A_P$ is proportional to $I^{16/7}$; using (25) and (38), Figure 10 highlights the substantial reduction in core dimensions in the proposed system, with a circulating current limited to 50% of the output current and a 0.1 $i_{circ}$ ripple current. Consequently, the proposed solution requires less than one-fifth of the magnetic core cross-section area compared to other designs.

According to Fig. 8(a), $i_{out}$ is evenly distributed across the two coupled inductors and consequently between the two battery modules in the proposed circuit and selected literature [15], [69, 72], [65-68, 70-71, 74]. Due to this benefit, two adjacent batteries can share the load, which reduces the peak load current of each module. Consequently, the RMS of the modules' current and then the conduction losses decreases. Furthermore, in Fig. 8(c), the load current is not shared between the modules in parallel, which increases the peak current of the batteries as well as the magnitude of the circulating currents among the modules and subsequently requires larger inductor core sizes.

Interestingly, only two examples from the literature [15] and the proposed circuit benefit from bidirectional energy transfer. The main advantages of the proposed topology and modulation scheme are significant reduction of the magnetic core size with identical energy transfer capability, bidirectional energy transfer with improved efficiency, better load sharing between modules, lower peak load current on the battery modules, and the possibility of combining modules with different voltages as well as characteristics [65], [69], [72].

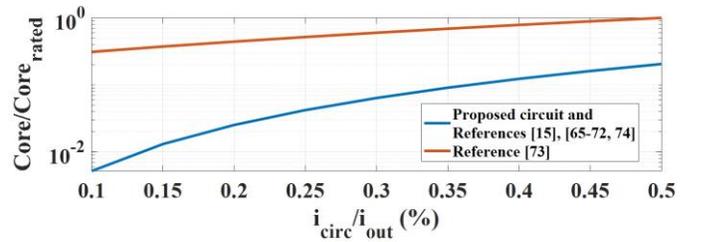

Fig. 10 Logarithmic comparison between core dimensions according to the ratio of $i_{circ}/i_{out}$.

### B. Comprehensive Comparison of Power Management Configurations: Efficiency, Complexity, and Flexibility

This section compares key elements of two power management systems configurations, specifically previous work that integrates batteries and ultracapacitors for microgrid applications [75-80] and the proposed circuit. Table III summarizes the comparison with a focus on the number of components, control complexity, flexibility, and system efficiency.

- Component count: Previous work [75-80] involves the use of separate bidirectional DC/DC converters for interfacing both batteries and ultracapacitors with the module. These converters require multiple power switches, inductors, and capacitors to manage bidirectional energy flow. Additionally, two converters are necessary—one for the battery and one for the ultracapacitor. This increases the overall component count and complexity of the system. In contrast, the proposed topology simplifies the architecture by using a quasi-DC/DC converter inherently available without adding extra components.

Table III
Comparison Between the Elements of Power Management Systems

| Converter topology | Interfacing elements | Energy transfer | Control complexity | Flexibility in energy management |
|---|---|---|---|---|
| [75] | DAB, C | bad | yes | no |
| [76] | DAB, C | bad | yes | no |
| [77] | DAB, C | bad | yes | no |
| [78] | DAB, C | bad | yes | no |
| [79] | DAB, C | bad | yes | no |
| [80] | DAB, C | bad | yes | no |
| Proposed converter | – | good | no | yes |

* Capacitor: C, Dual Active Bridge: DAB

- Control complexity: The use of dedicated DC/DC converters for batteries and ultracapacitors may provide flexibility but also tends to increase control complexity for dynamic load and energy sharing [75-80]. The presented system integrates the modes and requires fewer control variables. The coupled inductors further increase energy transfer efficiency and streamlines the control of energy flow between batteries of different voltages without the need for ultracapacitors.

- Flexibility in energy management: The combination of ultracapacitors for high power and batteries for energy in the prior art adds to the system's design complexity [75-80]. The proposed configuration demonstrates flexibility by dynamically managing energy between battery modules of different voltages, types, or even age. Coupled inductors and quasi-DC/DC conversion modes (buck, boost, and parallel) allow the system to rapidly distribute energy as needed during load fluctuations, such as EV acceleration or deceleration. The concept minimizes the inductance in the load path and maximizing the one for circulating currents to optimize energy transfer and reduce losses during high-power operation.

- Energy transfer efficiency: Although ultracapacitors are efficient in providing fast energy discharge for transient loads, the separate converters introduce power losses, particularly when switching between different operational modes. The use of coupled inductors in the MMSPC, in contrast, significantly increases energy transfer efficiency. The inductors minimize losses during energy exchange between battery modules due to their low resistance and magnetic losses. The system can further flexibly shift between buck, boost, and parallel without a mode change and minimize unnecessary multiple conversion (e.g., from batteries into capacitors and out of capacitors again into the load). This results in higher energy efficiency, especially during periods of fluctuating load demands.

## VI. SIMULATION RESULTS

*Scenario 1: Dynamic Output Power Above and Below Energy Module Capacity*

TABLE IV
SYSTEM PARAMETERS

| Parameter | Simulation | Experiment |
|---|---|---|
| Carrier frequency for quasi DC–DC circuits | 2 kHz | 2 kHz |
| Number of modules | 5 | 5 |
| Battery nominal voltage $v_{bn}$ | 22 V<$v_{b1}$<23 V<br>22 V<$v_{b2}$<23 V<br>22 V<$v_{b3}$<23 V<br>22 V<$v_{b4}$<23 V<br>22 V<$v_{b5}$<23 V | 22 V<$v_{b1}$<23 V<br>22 V<$v_{b2}$<23 V<br>22 V<$v_{b3}$<23 V<br>22 V<$v_{b4}$<23 V<br>22 V<$v_{b5}$<23 V |
| Coupled inductor | 25 μH | 25 μH |
| $L_{eff,CM}$ | 0 μH | ≈ 0 μH |
| $L_{eff,DM}$ | 100 μH | 100 μH |
| Switch | MOSFET<br>$r_{DS}$: 1 mΩ<br>$v_{DS}$: 100 V<br>$i_D$: 300 A<br>$t_r$: 20 ns<br>$t_f$: 20 ns | MOSFET<br>$r_{DS}$: 2 mΩ<br>$v_{DS}$: 100 V<br>$i_D$: 300 A<br>$t_r$: 13 ns<br>$t_f$: 17 ns |
| Load resistance | (1,10) Ω, 100 μH | (1,10) Ω, 100 μH |
| Output power | 0.2 – 2 kW | 0.2 – 2 kW |
| Output capacitance | 600 μF | 430 μF |
| DC inductance ($L_{filter}$) | 0.5 mH | 0.33 mH |

We modelled the proposed system and its modulation strategy in MATLAB/Simulink with a resistive-inductive load, as outlined in Table IV. The system uses a phase-shifted carrier (PSC) modulation at a 2 kHz carrier frequency and consists of five modules with voltage levels $v_{b1}$=22.7 V, $v_{b2}$=22.7 V, $v_{b3}$=22.4 V, $v_{b4}$=22.4 V, $v_{b5}$=22.4 V. Modules 1 and 2 act as energy modules for steady-state power, while Modules 3, 4, and 5 are power modules for high demand, such acceleration and regenerative braking.

Figure 11 displays the system's output voltage, current, and power. Until $t$ = 1 s, the output voltage reference is 70 V and the system maintains seven voltage levels with two energy modules and occasionally one power module to deliver 370 W of power. Each energy module contributes 300 W (Fig. 12) with a combined output of 600 W, which surpasses the load and recharges the power modules (Modules 3, 4, and 5 each absorb ~77 W, Figs. 13-15). The system's dc-dc mode and coupled inductors transfer the energy and balance load and charge.

At $t$ = 1 s, the output voltage increases to 105 V and eleven voltage levels with all five modules. The total output power rises to 815 W. The energy modules continuing to deliver 600 W together (Fig. 12). The power modules supply the remaining 215 W, approximately 71.5 W each. The difference in output between the energy modules stems from Module 2, which charges the power modules as it discharges slightly above 300 W. Module 2's battery current stays at –12.5 A, similar to the previous state, due to the energy transfer (Fig. 14). All power modules discharge evenly at about –4 A (Fig. 15). The current in the coupled inductors, reflects the continuous energy transfer from energy modules to power modules and vice versa (Fig. 16).

At 2 s, the load drops from 6 Ω to 2 Ω. The total output power requirement spikes to 2550 W. Figure 12 indicates that the energy modules maintain their contribution of 600 W (300 W each), while the power modules collectively provide an additional 1950 W. The DC/DC converter between Modules 2 and 3 enables energy modules to draw power from the power modules and stabilizes their output at exact 300 W each (Fig. 16. Despite the change in output power, the current in the energy modules remains stable at approximately –12.5 A, while the power module current rises to around –29 A (Figs. 14 and 15).

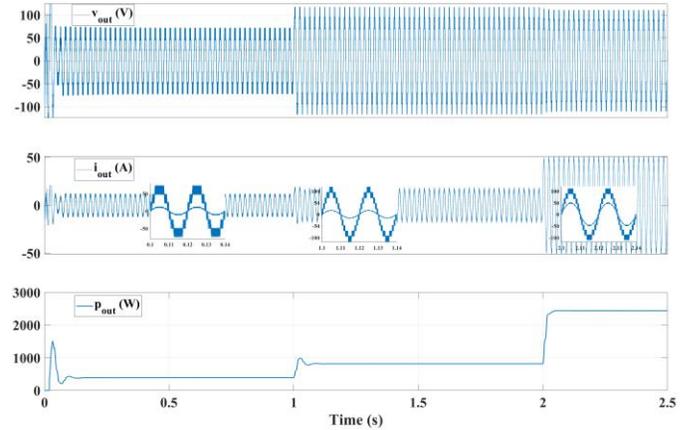

Fig. 11 Simulation results of $v_{out}$, $i_{out}$, and $p_{out}$.

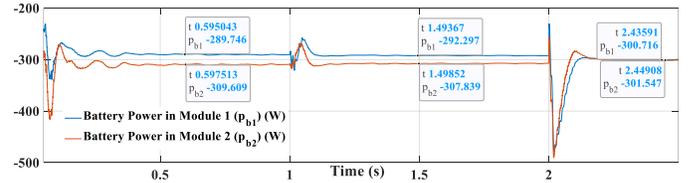

Fig. 12 $p_{b1}$ and $p_{b2}$ of the batteries in energy modules.

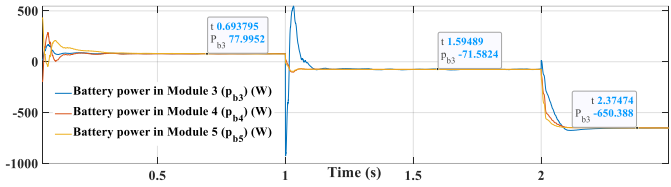
Fig. 13 $p_{b3}$, $p_{b4}$ and $p_{b5}$ of the batteries in power modules.

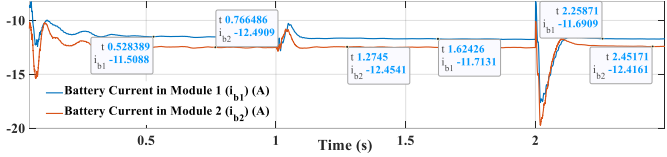
Fig. 14 $i_{b1}$ and $i_{b2}$ flowing through the batteries in energy modules.

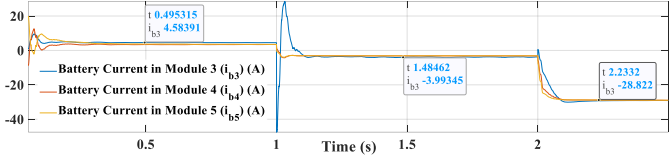
Fig. 15 $i_{b3}$, $i_{b4}$ and $i_{b5}$ flowing through the batteries in power modules.

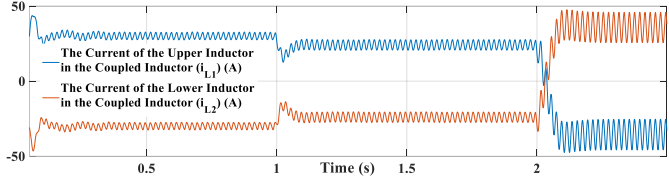
Fig. 16 $i_{L1}$, $i_{L2}$ currents between energy modules and power modules.

*Scenario 2: zero circulating current*

In this scenario, there is no active control for energy transfer between the energy and power modules. This means that the DC/DC and parallel modes between Modules 2 and 3 are not used, and the coupled inductor does not contribute to load balancing or charge distribution.

Figure 17 shows the system's output voltage, current, and power. Up to 0.5 s, with an output voltage set to 90 V, the system maintains nine voltage levels with four modules and a total output power of 815 W. Both energy and power modules deliver approximately 164 W each (Fig. 18). All modules maintain a uniform current of approximately –6.1 A. This consistency results from the absence of energy-sharing control. However, the parallel mode across all modules still ensures balanced charging. Accordingly, no current flows through the coupled inductors so that no energy is transferred (Fig. 20).

At 0.5 s, the output voltage changes from 90 V to 70 V and reduces the total power to 385 W. Energy and power modules contributes around 78 W each (Fig. 18) without any power exchange (Fig. 20). According to Equation (1), with $i_{circ} = 0$, the current through the coupled inductor equals half of $i_{out}$. Due to the parallel mode, the current in the energy and power module batteries is balanced (Fig. 19).

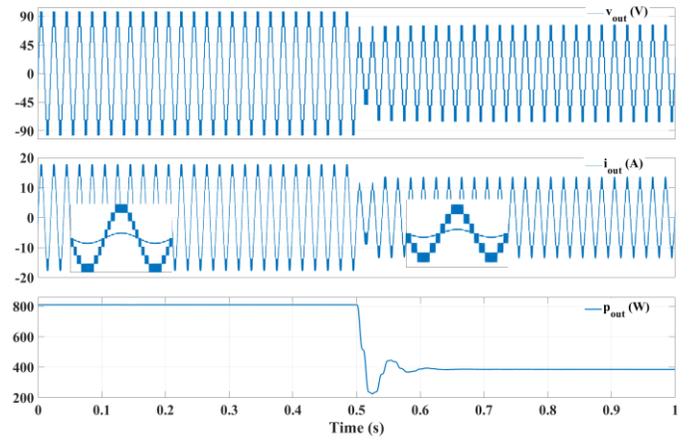
Fig. 17 Simulation results of $v_{out}$, $i_{out}$, and $p_{out}$ when $i_{ref (circ)}$ is zero.

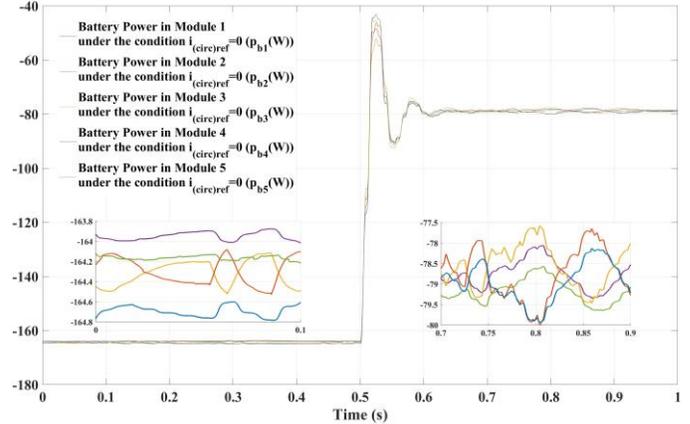
Fig. 18 $p_{b1}$, $p_{b2}$ $p_{b3}$, $p_{b4}$ and $p_{b5}$ of the batteries in energy and power modules.

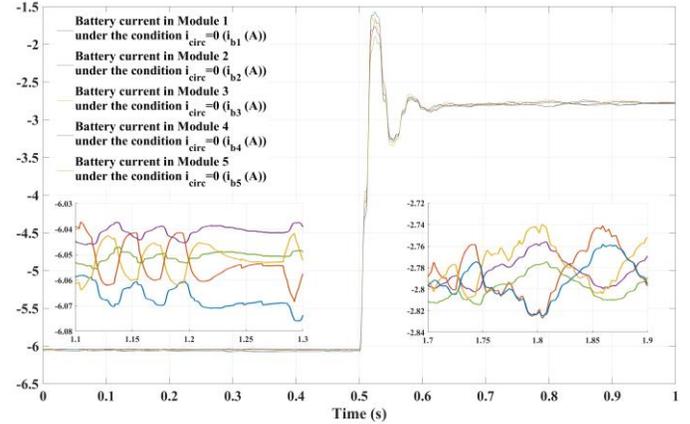
Fig. 19 $i_{b1}$, $i_{b2}$ $i_{b3}$, $i_{b4}$ and $i_{b5}$ of the batteries in energy and power modules.

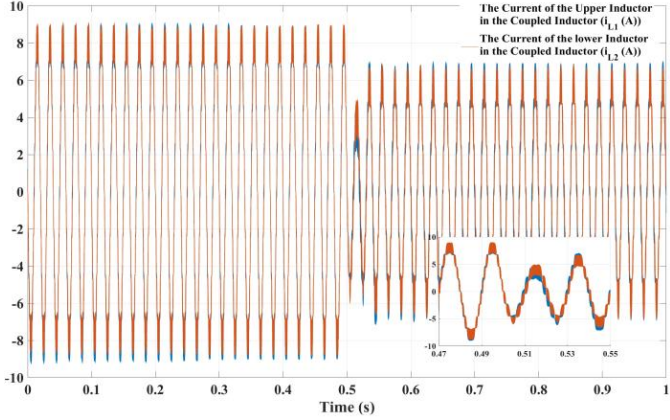

Fig. 20 $i_{L1}$, $i_{L2}$ currents between energy modules and power modules when $i_{circ} = 0$.

## VII. EXPERIMENTAL STUDY

We developed an experimental prototype (Fig. 21, system parameters per Table IV). As the simulation model, the experimental setup consists of five modules, and one coupled inductor between the groups of energy and power modules. Each module includes two full-bridge legs (IPT015N10N5 MOSFETs with an $r_{DS(on)} = 2$ mΩ). Each module contains a 6s2p lithium-ion battery, 22.2 V nominally, 5 Ah). To provide a low impedance path during switching transitions, ceramic capacitors (approximately 300 µF) is connected in parallel with each battery. LV 25-P and LA 55-P sensors (LEM) provide voltage and current measurements, which the controller (NI sbRIO 9627 with AMD Zynq 7020) samples at 10 kHz with 16-bit precision. The controller processes the output voltage reference ($v_{out-ref}$) and implements the control strategy of Fig. 7.

The coupled inductors use toroidal cores with part number B64290L0038X027. For comparison, if a conventional inductor were used instead to handle high current in parallel mode, Equation (27) would require an inductance value of 230 µH, rated for 10 A, considering parameters such as, $v_b^{max}$= 23 V, $m_{d2}$= 0.05, $f_{sw}$= 10 kHz, and $\Delta i_{diff} = 10\% \times 10$ A, The proposed solution, however, significantly reduces the current handled by the inductors to allow for a much smaller core. For testing purposes, the modules are set to have a voltage imbalance of around 5%, which, under conventional systems, would typically result in substantial circulating currents.

The topology is resilient to inductance tolerances as long as the self and mutual inductances of each coupled inductor within a group remain balanced. The measured inductance values for the coupled inductor are $L_1 = M_{12} = M_{21} = L_2 = 25$ µH ± 1 µH, yielding an approximate differential-mode inductance of 100 µH and a negligible common-mode inductance. This cancellation effect of self and mutual inductances ensures that the system behaves as intended even with slight variations.

*Scenario 1: Dynamic Output Power Above and Below Energy Module Capacity*

Figure 22 displays the system's output voltage, current, and power. Up to 0.5 s, with an output voltage reference of 70 V, the system maintains seven voltage levels, and uses two energy modules as well as occasionally one power module to deliver between 500 W to 600 W of output power. Each energy module contributes 300 W (Fig. 23) for a combined output of 600 W, which surpasses the load and recharges the power modules. Figures 24 and 25 indicate that Modules 3, 4, and 5 each absorb between 0 W to 30 W. The system's DC/DC mode and coupled inductors transfer energy to ensure balanced current flow for both load and charge balancing. The current in Module 2's battery is around –10.5 A (Fig. 23). The currents through all power module batteries align at approximately 2 A (Figs. 24 and 25).

At 0.5 s, the output voltage reference increases to 105 V and the system switches to eleven voltage levels with all five modules. The total output power rises to 1250 W – 1600 W. The energy modules continue to deliver 600 W. This power sustains the load and aids the power modules. Module 2 provides ~300 W as before (Fig. 23), while the power modules supply the remaining 650 W – 1000 W, approximately 230 W – 330 W each. The difference in output between the energy modules stems from Module 2, which charges the power modules and itself discharges at about 300 W. Module 2's battery current remains at –10.5 A due to the energy transfer. All power modules discharge evenly at about –13 A to –15 A (Figs. 24 and 25).

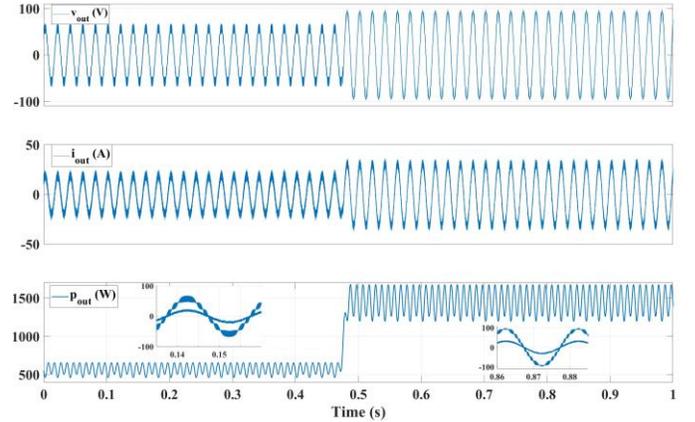

Fig. 22 Simulation results of $v_{out}$, $i_{out}$, and $p_{out}$.

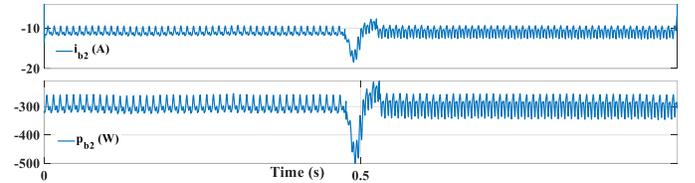

Fig. 23 $i_{b2}$ and $p_{b2}$ of the batteries in energy modules.

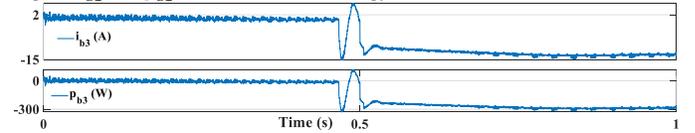

Fig. 24 $i_{b3}$ and $p_{b3}$ of the batteries in energy modules.

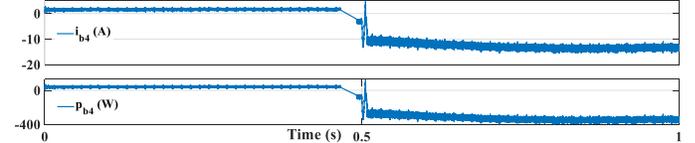

Fig. 25 $i_{b4}$ and $p_{b4}$ of the batteries in energy modules.

*Scenario 2: Output Voltage and Current Control with Load Change*

The experimental results validate the system's performance under various loads. Initially, the output voltage and current are both stable and sinusoidal at a power factor of nearly 1 (Fig. 26). As the load changes to a dynamic resistive configuration, the voltage remains stable, and the current closely tracks the voltage to maintain a high power factor.

When the load shifts to inductive (Fig. 27), the system continues to deliver consistent voltage and current waveforms. Despite a small phase shift, the power factor remains high and the output quality high.

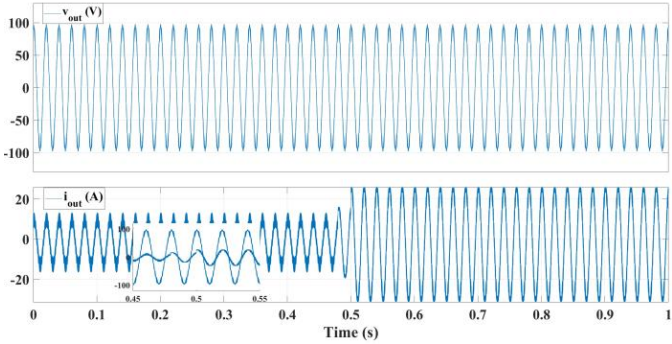

Fig. 26 Experimental results of dynamic changes of $v_{out}$, and $i_{out}$ in resistive load.

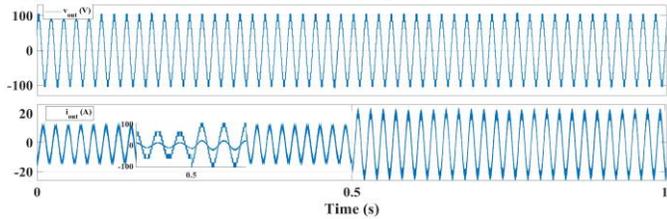

Fig. 27 Experimental results of dynamic changes of $v_{out}$, and $i_{out}$ in inductive load.

### Scenario 3: Long-Term Current and Voltage Control with Varying Module Voltages—No Load Exchange, Load Transfer, and Reversal

In this scenario, the system is tested for its ability to manage energy transfer between modules with different voltages when the load is maintained at around 850 W. Initially, there is no energy exchange between the modules, which allows the system to reach a steady-state condition. The starting voltages are slightly different across the modules and create an intentional imbalance to observe the controller's response.

The modulation strategy forces a transfer of load from one side to the other (Figs. 28 and 29). Voltage $v_{b2}$ initially stabilizes at around 23.86 V. Current $i_{b2}$ features minor fluctuations but generally trends towards a steady-state value with positive slope as the system adjusts. For Module 3, $v_{b3}$ starts at approximately 23 V (Fig. 29). Over time, the load is shifted from Module 2 to Module 3 and causes a slight drop in $v_{b3}$, while $v_{b2}$ remains relatively stable.

This controlled transfer of energy demonstrates the effectiveness of the DC/DC mode in balancing and distributing the load between modules even under unbalanced conditions. Module 3 supplies power to the load and charges Module 2. Figure 30 illustrates the current in the coupled inductors. The currents in the upper ($i_{L1}$), and lower ($i_{L2}$) branches are balanced. As the load transfer begins, a distinct difference in current magnitude is observed and indicates energy flow from power modules to energy modules. This matches the behavior expected from the control strategy and aligns well with Equation (1), which shows

$$i_{L1} + i_{L2} = i_{out}, \tag{39}$$

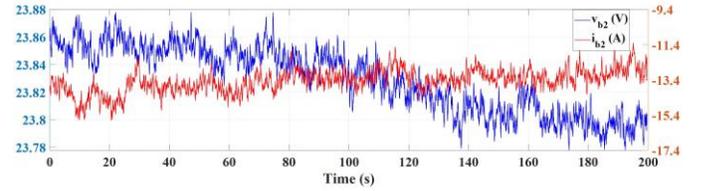

Fig. 28 Experimental results of static changes of $v_{b2}$, and $i_{b2}$ in inductive load in static condition when it is charged by the power modules.

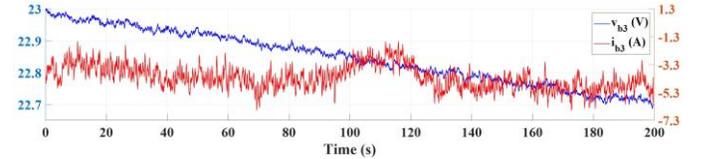

Fig. 29 Experimental results of static changes of $v_{b3}$, and $i_{b3}$ in inductive load in static condition when it is discharged in the energy modules.

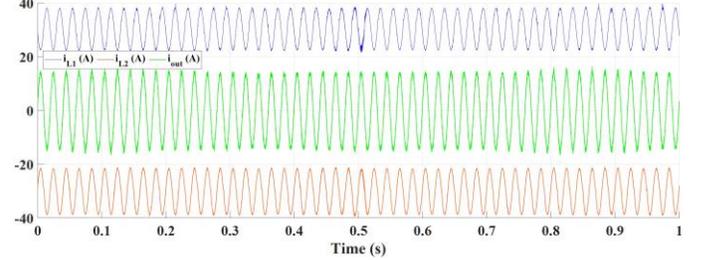

Fig. 30 $i_{L1}$, $i_{L2}$, $i_{out}$ currents between energy modules and power modules in static condition when the power modules are charging the energy modules.

In this part of the experiment, the direction of energy transfer is reversed to charge the power modules from the energy modules and test the control strategy's effectiveness in handling dynamic energy flow while maintaining system stability.

During the $m_{d2}(+)$ mode (Fig. 31), Module 2 is required to discharge slightly more than before; $i_{b2}$ decreases to around –14 A instead of –13 A. This higher discharge current occurs because Module 2 needs to supply additional energy to charge Module 3, while it also proves power to the load. Although both modules' voltages decrease over time due to continuous power delivery, the discharge rate of Module 3 is slower than that of Module 2. This behavior indicates that Module 3 is receiving energy support from Module 2 and effectively balances the power demand.

Figure 32 graphs $i_{b3}$ and $v_{b3}$. The voltage remains relatively stable initially, but as charging progresses, $i_{b3}$ shows a distinct increase in current, which reaches approximately –3 A. The negative currnet confirms that Module 3 is charged by the energy modules, while they concurrently support the load.

Figure 33 illustrates the current profiles for $i_{L1}$, $i_{L2}$, and $i_{out}$ during this reversed energy flow scenario. According to Equation (1), the direction of $i_{circ}$ is now reversed, resulting in a change in polarity for $i_{L1}$ and $i_{L2}$. In contrast to the previous mode $m_{d2}(-)$, where $i_{L1}$ was positive, it now becomes negative; the negative value reflects the energy transfer from the energy modules to the power modules. Despite this reversal, the total output current $i_{out}$ remains consistent and confirms the expected energy flow as per the control strategy.

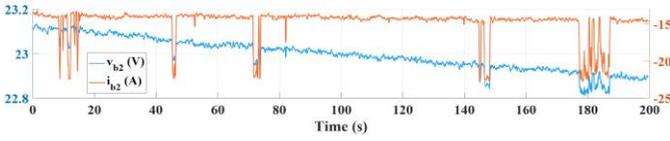

Fig. 31 Experimental results of static changes of $v_{b2}$, and $i_{b2}$ in inductive load in static condition when it is discharged in the power modules.

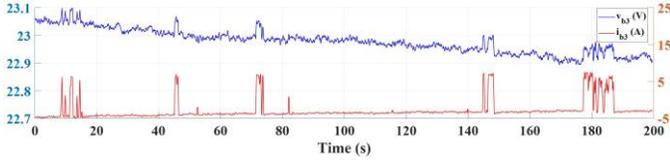

Fig. 32 Experimental results of static changes of $v_{b3}$, and $i_{b3}$ in inductive load in static condition when it is charged by the energy modules.

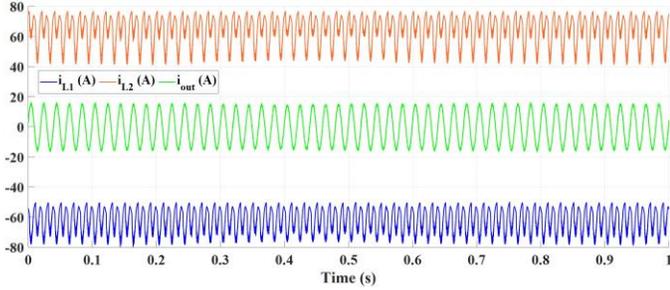

Fig. 33 $i_{L1}$, $i_{L2}$, $i_{out}$ currents between energy modules and power modules in static condition when the energy modules are charging the power modules.

In the last phase of this scenario, the reference circulating current $i_{\text{circ(ref)}}$ is set to zero to effectively disable active energy transfer between the modules. The voltages of both batteries start from an equal value of 23.9 V, and their currents remain relatively stable throughout the test (Figs. 34 and 35). This stability indicates that no energy is exchanged between the energy and power modules.

Figure 36 further validates this condition by illustrating that $i_{\text{circ}}$ is maintained at zero. Additionally, $i_{\text{out}}$ is approximately twice the value of $i_{L1}$ and $i_{L2}$. This relationship aligns with the expected behavior when $i_{\text{circ}} = 0$ and confirms that the energy modules operate independently without contributing to inter-module energy transfer.

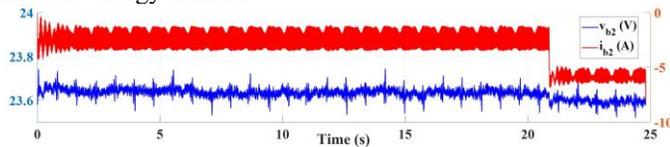

Fig. 34 Experimental results of dynamic changes of $v_{b2}$, and $i_{b2}$ in inductive load when $i_{\text{ref(circ)}}$ is zero.

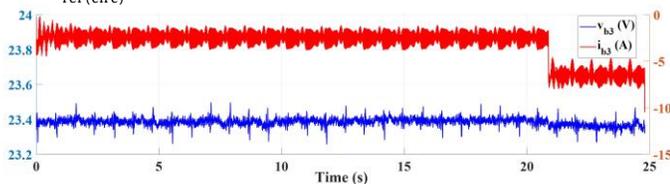

Fig. 35 Experimental results of dynamic changes of $v_{b3}$, and $i_{b3}$ in inductive load when $i_{\text{ref(circ)}}$ is zero.

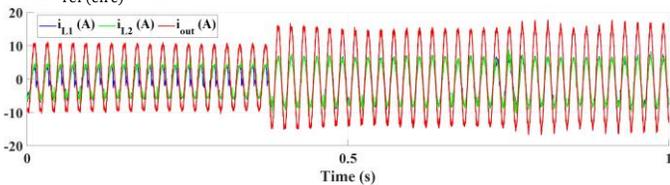

Fig. 36 Dynamic $i_{L1}, i_{L2}, i_{out}$ currents between energy modules and power modules when $i_{\text{ref(circ)}}$ is zero.

## VIII. CONCLUSION

This paper introduced a novel reconfigurable series/parallel battery for better dynamic power management in reconfigurable battery systems, such as electric vehicles. The proposed architecture combines energy and power modules connected through coupled inductors to manage energy transfer and load balancing across the energy and power modules. Simulations and experimental results demonstrate consistent performance in various scenarios, including dynamic power changes, zero circulating current, and adaptive control under voltage imbalance.

The reconfigurable battery demonstrates significant advantages over other approaches that combine energy and power modules; it reduces the component count, simplifies control, and increases energy efficiency. The integrated quasi-DC/DC modes enabled smooth transitions between buck, boost, and parallel operation without additional hardware for low losses and small magnetic core size.